# DESIGN OF POWER SYSTEM STABILIZER USING FUZZY BASED SLIDING MODE CONTROL TECHNIQUE

LATHA.R
*Department of Instrumentation and Control Systems Engineering, PSG College of Technology, Coimbatore, 641004, India.*
*rla@ice.psgtech.ac.in*

KANTHALAKSHMI.S
*Department of Instrumentation and Control Systems Engineering, PSG College of Technology, Coimbatore, 641004, India.*
*klakshmiramesh@yahoo.co.in, skl@ice.psgtech.ac.in*

KANAGARAJ.J
*Department of Electrical and Electronics Engineering, PSG College of Technology, Coimbatore, 641004, India*
*jkk@eee.psgtech.ac.in*

Power systems are usually large non-linear systems, which are often subjected to low frequency electromechanical oscillations. Power System Stabilizers (PSSs) are often used as effective and economic means for damping the generators' electromechanical oscillations and enhance the overall stability of power systems. Power system stabilizers have been applied for several decades in utilities and they can extend power transfer stability limits by adding modulation signal through excitation control system. Sliding mode control is one of the main methods employed to overcome the uncertainty of the system. This controller can be applied very well in presence of both parameter uncertainties and unknown nonlinear function such as disturbance. To enhance stability and improve dynamic response of the system operating in faulty conditions a Fuzzy based Sliding Mode Control PSS is developed for a multimachine system with two generators and it is compared with the Conventional PSS, Fuzzy based PSS, Sliding Mode based PSS.

## 1. INTRODUCTION

With the increase in the demand for electrical power and the need to operate power systems closer to their limits of stability, at a faster and more flexible manner in the deregulated competitive environment, modern power systems has reached a stressed conditions. These cause unstable or poorly damped oscillations that are observed more often in today's power systems. The instability problems caused by low frequency inter-area oscillations that occur due to weak interconnected power systems are therefore becoming significant. Controlling the voltage throughout the network, as well as damping power frequency oscillations presents a continuous challenge to power system engineers.

In recent years, considerable efforts have been made to enhance the dynamic stability of power systems. In order to reduce the undesirable oscillatory effect and improve the system dynamic performance, supplementary signal are introduced as a measure to increase the damping. One of the cost effective approach is fitting the generators with a feedback controller to inject a supplementary signal at the voltage reference input of the automatic voltage regulator to damp the oscillations. This device known as a Power System Stabilizer (PSS). [1-2] They provide good damping; thereby contribute in stability enhancement of the power systems.

Over the past four decades, various control methods have been proposed for PSS design to improve overall system performance. Because of their simple structure, flexibility and ease of implementation, conventional PSS of the lead-lag compensation type have been adopted by most utility companies. But the performance of these stabilizers can be considerably degraded with the changes in the operating condition during normal operation. Since power systems are highly nonlinear, conventional fixed-parameter PSSs cannot cope with great changes in the operating conditions.

## 2. MODELLING OF PLANT

The modelling of a simple 500 KV transmission system containing two hydraulic power plants is shown in Fig. 1. PSSs are used to improve transient stability and power system oscillations damping. Despite the simple structure of the illustrated power system in the figure, the phasor simulation method can be used to simulate more complex power grids.

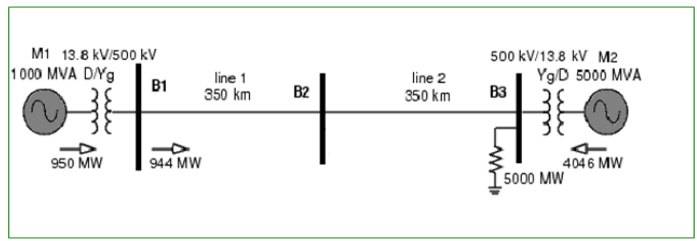

Fig. 1. Single line diagram of power system

A 1000 MVA hydraulic generation plant (M1) is connected to a load centre through a long 500 KV, 700 km transmission line. A 5000 MW of resistive load is modelled as the load centre. The remote 1000 MVA plant and a local generation of 5000 MVA (plant M2) feed the load.

A load flow has been performed on this system with plant M1 generating 950 MW so that plant M2 produces 4046 MW. The line carries 944 MW which is close to its surge impedance loading (SIL = 977 MW). The two machines are equipped with a hydraulic turbine and governor (HTG), excitation system, and power system stabilizer (PSS).[3]

## 3. POWER SYSTEM STABILIZER

The generic Power System Stabilizer (PSS) block is used in the model to add damping to the rotor oscillations of the synchronous machine by controlling its excitation current. Any disturbances that occur in power systems can result in inducing electromechanical oscillations of the electrical generators. Such oscillating swings must be effectively damped to maintain the system stability and reduce the risk of outage. The output signal of the PSS is used as an additional input ($V_{stab}$) to the excitation system block. The PSS input signal can be either the machine speed deviation ($d\omega$) or its acceleration power,

$P_a = P_m - P_e$ (difference between the mechanical power and the electrical power). Fig.2. shows Generic PSS. [3]

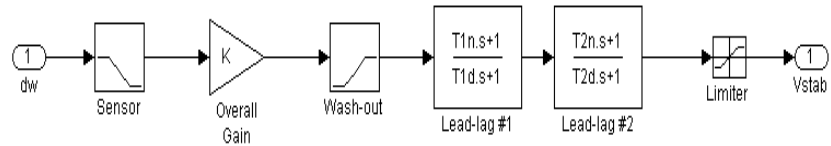

Fig. 2. Generic power system stabilizer

To ensure a robust damping, a moderate phase advance has to be provided by the PSS at the frequencies of interest in order to compensate for the inherent lag between the field excitation and the electrical torque induced by the PSS action. The model consists of a low-pass filter, a general gain, a washout high-pass filter, a phase-compensation system, and an output limiter. The general gain (K) determines the amount of damping produced by the stabilizer. The washout high-pass filter eliminates low frequencies that are present in the dω signal and allows the PSS to respond only to speed changes. The phase-compensation system is represented by a cascade of the two first-order lead-lag transfer functions used to compensate the phase lag between the excitation voltage and the electrical torque of the synchronous machine. [4,5].

## 4. FUZZY LOGIC CONTROL

Fuzzy Logic Controls (FLCs) are very useful when an exact mathematical model of the plant is not available however; experienced human operators are available for providing qualitative rules to control. The essential part of the fuzzy logic controller is a set of linguistic control rules related by dual concepts of fuzzy implication and the compositional rule of inference. The fuzzy logic controller is simpler and fastest methodology. It does not need any exact system mathematical model and it can handle nonlinearity of arbitrary complexity. It is based on the linguistic rules with an IF-THEN general structure, which is the basis of human logic. The structure of fuzzy controller is shown in Fig 3. It consists of fuzzification inference engine and defuzzification blocks. [6]

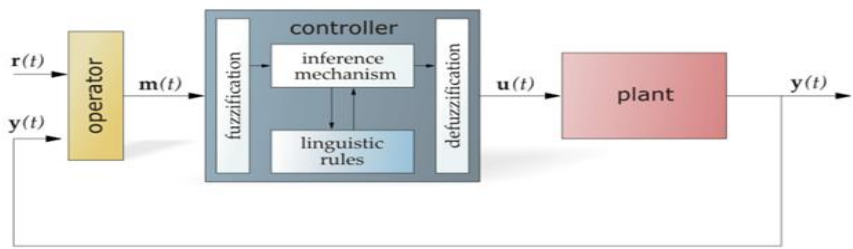

Fig. 3. Structure of fuzzy controller

Here the, input variables are change in speed deviation (dω) and change in acceleration (da) and the output variable is stabilizing voltage ($V_{stab}$). The membership functions for dω, da, and Vstab are as shown below in Fig. 5, Fig. 6 and Fig. 7 respectively. Table 1 shows the rule base for fuzzy controller.

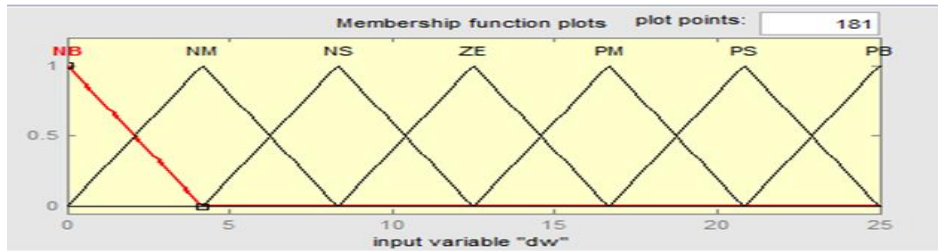

Fig. 4. Membership function plot of speed deviation (dω)

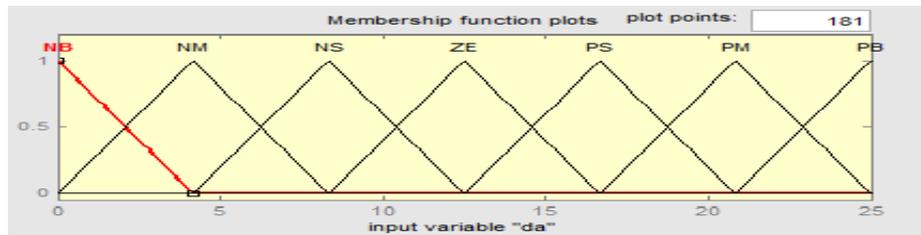

Fig. 5. Membership function plot of acceleration (da)

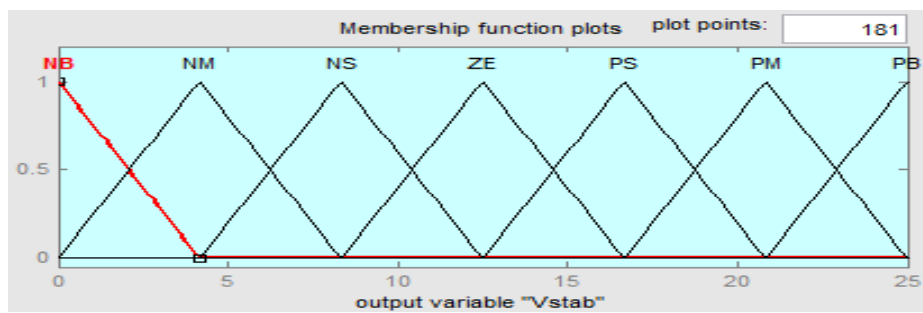

Fig. 6. Membership function plot of stabilizing voltage ($V_{stab}$)

TABLE 1 RULE BASE FOR FUZZY CONTROLLER

| Speed Deviation | Acceleration | | | | | | |
|---|---|---|---|---|---|---|---|
| | NB | NM | NS | ZE | PS | PM | PB |
| NB | NB | NB | NB | NB | NM | NM | NS |
| NM | NB | NM | NM | NM | NS | NS | ZE |
| NS | NM | NM | NS | NS | ZE | ZE | PS |
| ZE | NM | NS | NS | ZE | PS | PS | PM |
| PS | NS | ZE | ZE | PS | PS | PM | PM |
| PM | ZE | PS | PS | PM | PM | PM | PB |
| PB | PS | PM | PM | PB | PB | PB | PB |

## 5. SLIDING MODE CONTROL

Sliding Mode Control (SMC) is one of the main methods employed to overcome the uncertainty of the system. This controller can be applied in presence of both parameters uncertainties and unknown nonlinear function such as disturbance. Sliding mode control technique has been used to control robots, motors, mechanical systems, etc. and assure the desired behavior of closed loop system. [7]

### 5.1 Dynamic Model of Synchronous Generator

Consider a multi-machine power system consisting of two synchronous generators with load as shown in Fig. 7. This system is a two area power system, the two areas are identical and each include a generator equipped with fast acting excitation systems.

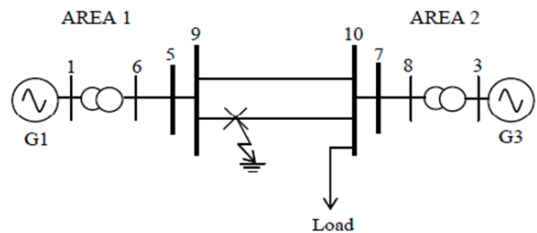

Fig. 7. Multimachine system with two generators

The equations describing a third order model of a synchronous generator (for jth generator) is given as:

$$\dot{\delta}_j = w_j(t) - w_{oj}$$
$$\dot{w}_j(t) = \frac{K_{Dj}}{2H_j}(w_j(t)-w_{oj}) + \frac{w_{oj}}{2H_j}(P_{mj}-P_{ej}(t)) \tag{1}$$
$$\dot{E}'_{qj}(t) = \frac{1}{T'_{doj}}(E_{Fj}(t) - E_{qj}(t))$$

where

$$E_q(t) = \frac{X_{dj}}{X'_{dj}}E'_{qj}(t) - \frac{x_{dj}-x'_{dj}}{X'_d}V_s \cos(\delta_j(t))$$
$$E_{Fj(t)} = k_{cj}u_{Fj}(t) \tag{2}$$
$$P_{ej}(t) = \frac{V_s E_{qj}(t)}{X_d}\sin(\delta_j(t))$$

and $\delta_j(t)$ is the rotor angle of the jth generator (radians),
$w_j(t)$ is the speed of the rotor of the jth generator (radian/sec),
$w_{oj}$ is the synchronous machine speed of the jth generator (radian/sec),
$K_{Dj}$ is the damping constant of the jth generator (pu),
$H_j$ is the inertia constant of the jth generator (sec),
$P_{mj}$ is the mechanical input power of the jth generator (pu),
$P_{ej}(t)$ is the active electrical power delivered by the jth generator (pu),
$E_{qj}(t)$ is the EMF of the q-axis of the jth generator (pu),
$E'_{qj}(t)$ is the transient EMF in the q-axis of the jth generator (pu),
$E_{Fj}(t)$ is the equivalent EMF in the excitation winding of the jth generator (pu),
$T'_{doj}$ is the d-axis transient short circuit time constant of the jth generator (sec),
$k_{Cj}$ is the gain of the excitation amplifier of the jth generator,
$u_{fj}(t)$ is the control input of the excitation amplifier with gain $k_{Cj}$,
$x'_{dj}$ is the d-axis transient reactance of the jth generator (pu),
$X_{dj}$ is the total direct reactance of the system (pu),
$X'_{dj}$ is the total transient reactance of the system (pu),
$V_s$ is the infinite bus voltage (pu) and j denotes the no. of generator.
The states of the system for jth generator is as follows:
$$x_{1j}(t) = \delta_j(t)$$
$$x_{2j}(t) = w_j(t)-w_{oj} \tag{3}$$
$$x_{3j}(t) = E'_{qj}(t)$$

Hence state vector for each generator will be:
$$x_{1j}(t) = [x_{1j}(t)\ x_{2j}(t)\ x_{3j}(t)]^T \tag{4}$$

Also the control input $u_j(t)$ is taken to be:
$$u_j(t) = \frac{k_c(j)}{T'_{doj}}u_{fj}(t) \tag{5}$$

The nonlinear equations of the system, define the following constants for each generator:
$$\alpha_{1j} = \frac{k_{cj}}{2H_j}$$
$$\alpha_{2j} = \frac{w_{oj}}{2H_j X'_{dj}}V_s$$

$$\alpha_{3j} = \frac{w_{oj}\ (x_{dj} - x'_{dj})}{4\ H_j\ X_{dj}\ X'_{dj}} V_s^2$$

$$\alpha_{4j} = \frac{w_{oj}}{2H_j} P_{mj}$$

$$\alpha_{5j} = \frac{-1}{T'_{doj}} \frac{X_{dj}}{X'_{dj}}$$

$$\alpha_{6j} = \frac{x_{dj} - x'_{dj}}{T'_{doj}\ X'_{dj}} V_s \tag{6}$$

Substituting (6) in (1) and (2), we get set of equations describing jth generator as given below.

$$\dot{x}_1 = x_{2j}(t)$$
$$\dot{x}_2 = \alpha_{1j} x_{21}(t) + \alpha_{2j} x_{3j}(t) \sin\left(\alpha_{1j} x_{21}(t)\right) + \alpha_{3j} \sin\left(2x_{1j}(t)\right) + \alpha_{4j}$$
$$\dot{x}_3 = \alpha_{5j} x_{3j}(t) + \alpha_{6j} \cos\left(x_{1j}(t)\right) + u_j(t) \tag{7}$$

The desired values of the system state vector for the jth generator is

$$x_{Dj} = [x_{1dj}\ x_{2dj}\ x_{3dj}] \tag{8}$$

where $x_{1dj}\ x_{2dj}\ x_{3dj}$ are the desired value of state.

The control input which enables the system to achieve the desired states is denoted by $u_{dj}$.

The deviations of the rotor angle of each generator from its desired value are taken as output of each system.

$$y_j(t) = x_{1j}(t) - x_{1dj} \tag{9}$$

The desired values are calculated by equating $\dot{x}_1, \dot{x}_2, \dot{x}_3$ to zero. The values of $x_{1dj}\ x_{2dj}$ and $x_{3dj}$ are derived as follows:

$$\left(-\frac{\alpha_{1j}\alpha_{6j}}{2\alpha_{5j}} + \alpha_{3j}\right) \sin(2x_{1dj}) - \frac{\alpha_{2j}}{\alpha_{5j}} u_{dj} \sin(x_{1dj}) + \alpha_{4j} = 0$$
$$x_{2dj} = 0$$
$$x_{3dj} = -\frac{\alpha_{6j}}{\alpha_{5j}} \cos(x_{1dj}) - \frac{1}{\alpha_{5j}} u_{dj} \tag{10}$$

### 5.2 Controller Design

The objective of this section is to design a controller based on sliding mode theory for synchronous generator so that it regulates the states of the system to their desired values and maintain the stability of the system in operation point in the presence of uncertainty and also increase the rate of oscillation damping. The equations (7) and (9) describing the synchronous generator are highly nonlinear.

Therefore, in first step, to facilitation design of nonlinear controller for each generator, a change of variable $z_j(t)$ is considered, such that:

$$z_{1j}(t) = x_{1j}(t) - x_{1dj}$$
$$z_{2j}(t) = x_{2j}(t)$$
$$z_{3j}(t) = \alpha_{1j} x_{21}(t) + \alpha_{2j} x_{3j}(t) \sin\left(\alpha_{1j} x_{21}(t)\right) + \alpha_{3j} \sin\left(2x_{1j}(t)\right) + \alpha_{4j} \tag{11}$$

From (10) and (11) it is obvious that if $z_j(t)$ converges to zero as $t \to \infty$, then $x_j(t)$ converges to as $t \to \infty$. The inverse of the transmission given in (11) is

$$x_{1j}(t) = z_{1j}(t) - x_{1dj}$$
$$x_{2j}(t) = z_{2j}(t)$$

$$x_{3j}(t) = \frac{-\alpha_{1j}x_{21}(t) - \alpha_{3j}\sin(2x_{1j}(t)) - \alpha_{4j}}{\alpha_{2j}\sin(\alpha_{1j}x_{21}(t))} \tag{12}$$

Substituting (7) in (11), the equations of the synchronous generator can be written as function of the new variable such that

$$\dot{z}_{1j}(t) = z_{2j}(t)$$
$$\dot{z}_{2j}(t) = z_{3j}(t)$$
$$\dot{z}_{3j}(t) = f_j(z) + G_j(z) u_j$$
$$y_j(t) = z_{1j}(t) \tag{13}$$

Where:

$$f_j(z) = ((\alpha_{1j} + \alpha_{5j}) z_{3j} - \alpha_{1j}\alpha_{5j}z_{2j} + \left(\frac{1}{2}\alpha_{1j}\alpha_{6j} - \alpha_{3j}\alpha_{5j}\right)\sin(2(z_{1j} + x_{1dj}))) + 2\alpha_{3j}z_{2j}\cos\left(2(z_{1j} + x_{1dj})\right) + z_{2j}\cot(z_{1j} + x_{1dj})\left(z_{3j} - \alpha_{1j}z_{2j} - \alpha_{3j}\sin\left(2(z_{1j} + x_{1dj})\right) - \alpha_{4j}\right) - \alpha_{4j}\alpha_{5j} \tag{14}$$

and $G_j(z) = \alpha_{2j}\sin(z_{1j} + x_{1dj})$ \hfill (15)

In the original coordinate, the functions $f_j(z) = f_{j1}(x)$ and $G_j(z) = G_{j1}(x)$ are:

$$f_{j1}(x) = \alpha_{1j}(\alpha_{1j}x_{2j} + \alpha_{2j}x_{3j}\sin(x_1) + \alpha_{3j}\sin(2x_1) + \alpha_{4j}) + \alpha_{2j}(\alpha_{5j}x_{3j} + \alpha_{6j}\cos(x_1))\sin(x_{1j}) + \alpha_{2j}x_{2j}x_{3j}\cos(x_{1j}) + 2\alpha_{3j}x_{3j}\cos(2x_{1j}) \tag{16}$$

and $G_{1j}(x) = \alpha_{2j}\sin(x_{1j})$ \hfill (17)

The model of the synchronous generator given by (14) will be used for designing the sliding mode controller. Then the designed controller will be transformed into the original coordinate as given in (12).

*5.2.1 Design of Sliding Surface*

The second step of the SMC design process is the design of the sliding surface. The sliding surface for each generator is as follows:

$$S = \ddot{y}_j + \rho_{1j}\dot{y}_j + \rho_{1j}y_j = z_{3j} + \rho_{1j}z_{2j} + \rho_{2j}z_{1j} \tag{18}$$

where coefficients $\rho_{1j}$ and $\rho_{2j}$ are positive scalars and are chosen to obtain the desired transient response of the output of the system. The switching surface can be written as a function of $x_{1j}(t), x_{2j}(t)$, and $x_{3j}(t)$ such that:

$$S_j = \alpha_{1j}x_{2j} + \alpha_{2j}x_{3j}\sin x_{1j} + \alpha_{3j}\sin 2x_{1j} + \alpha_{4j} + \rho_{1j}x_{2j} + \rho_{2j}(x_{1j} - x1_{dj}) \tag{19}$$

The choice of the switching surface guarantees that the output of the system converges to zero as $t \to \infty$ on the sliding surface $S_j(x) = 0$.

The third step of the proposed sliding mode controller process is to design the control function that provides the motion on the sliding surface, such that:

$$u_j(t) = \frac{-1}{G_j(z)}(f_j(z) + \rho_{1j}z_{3j} + \rho_{2j}z_{2j} + \eta_j \, \text{sign}(z_{3j} + \rho_{1j}z_{2j} + \rho_{2j}z_{1j})) \tag{20}$$

where $\eta_j$ is a positive scalar and is determined by designer.

On differentiating equation (19) with respect to the time, $\dot{S}$ is obtained as follows

$$\dot{S} = \dddot{y}_j + \rho_{1j}\ddot{y}_j + \rho_{2j}\dot{y}_j$$
$$= f_j(z) + G_j(z)u_j + \rho_{1j}z_{3j} + \rho_{2j}z_{2j} \tag{21}$$

Substituting (20) in (21)

$$\dot{S} = f_j(z) + \rho_{1j}z_{3j} + \rho_{2j}z_{2j} + (-f_{1j}(z) - \rho_{1j}z_{3j} + \rho_{2j}z_{2j} - \eta_j \, sign(z_{3j} + \rho_{1j}z_{2j} + \rho_{2j}z_{1j}))$$
$$= -\eta_j \, sign\left(z_{3j} + \rho_{1j}z_{2j} + \rho_{2j}z_{1j}\right) = -\eta_j \, sign(S) \tag{22}$$

Hence
$$S_j\dot{S}_j = -S_j\eta_j \, sign(S_j) = -\eta_j|\dot{S}| \tag{23}$$

Therefore the dynamics of $S_j$ in (23) guarantees that $S_j\dot{S}_j < 0$. The proposed sliding mode controller guarantees the asymptotic convergence of $z_j(t)$ to zero as $t \to \infty$.

Substituting (16), the controller function given in (20) can be written in the original coordinate as follow:

$$u_j = \frac{1}{\alpha_{2j}\,sin\,x_{1j}} \left[ \begin{array}{c} \left(-\alpha_{1j} - \rho_{1j}\right)\left(-\alpha_{1j}x_{2j} + \alpha_{2j}x_{3j}\,sin\,x_{1j} + \alpha_{3j}\,sin\,2x_{1j} + \alpha_{4j}\right) + \\ \left(-\alpha_{2j}(\alpha_{5j}x_{3j} + \alpha_{6j}\,cos\,x_{1j}\,sin\,x_{1j} - \alpha_{2j}x_{2j}x_{3j}\,cos\,x_{1j} - 2\alpha_{3j}x_{2j}\,cos\,2x_{1j}\right) + \\ \left(-\rho_{2j}x_{2j} - \eta_j sign(S_j)\right) \end{array} \right]$$

(24)

where
$$S_j = \alpha_{1j}x_{2j} + \alpha_{2j}x_{3j}\,sin\,x_{1j} + \alpha_{3j}\,sin\,2x_{1j} + \alpha_{4j} + \rho_{1j}x_{2j} + \rho_{2j}(x_{1j} - x1_{dj}) \tag{25}$$

The control signal $u_j$ is stabilizing voltage.

## 6. FUZZY BASED SLIDING MODE CONTROL

The Fuzzy Sliding Mode Control (FSMC) technique, is an integration of variable structure control and FLC, provides a simple way to design FLC systematically. The main advantage of FSMC is that the control method achieves asymptotic stability of the system. Another feature is that the method can minimize the set of FLC and provide robustness against model uncertainties and external disturbances. In addition, the method is capable of handling the chattering problem that arises in traditional SMC.

The conventional sliding mode control will make the system converge to the sliding surface at a rate proportional to $\eta_j$, however on convergence to the surface the chattering present in the system would also be proportional to the $\eta_j$. In order to get maximum advantage of the controller structure, $\eta_j$ should be tuned to be of high magnitude when the state is approaching the sliding surface, thus helping for faster convergence and when the state is in sliding mode, $\eta_j$ should be tuned to be of low magnitude in order to reduce chattering. This can be tuned by fuzzy controller. [8]

### *6.1 Design of Fuzzy based SMC*

The Fuzzy based SMC-PSS consists a single - input, a single - output component. The input variable is considered as Sliding surface 'S' and the output variable is considered as tuning parameter ′$\eta_j'$′. The fuzzy inference mechanism contains seven rules. The membership function for Sliding surface ′S′ is as shown in Fig. 8. The membership function for tuning parameter ′$\eta_j$′ is as shown in Fig. 9. The Table 2 gives the rule base for the fuzzy based SMC controller.

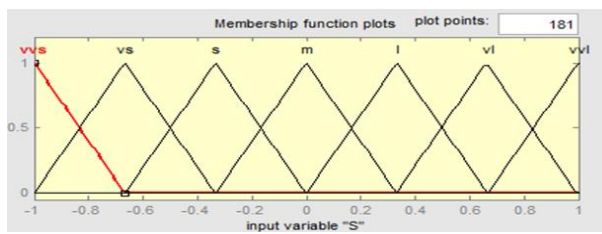
Fig.8. Membership function plot of sliding surface (S)

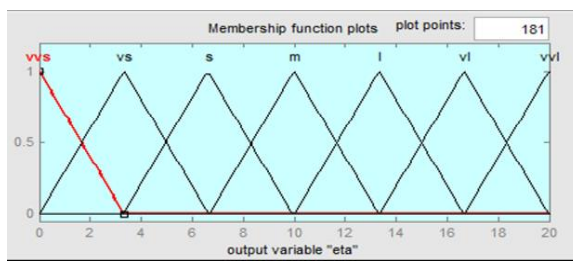
Fig.9. Membership function plot of tuning parameter $(\eta_j)$

TABLE 2. RULE BASE FUZZY BASED SMC

| Rule No. | Switching surface $|s|$ | Tuning parameter $(\eta_j)$ |
|---|---|---|
| 1 | VVS | VVS |
| 2 | VS | VSS |
| 3 | S | S |
| 4 | M | M |
| 5 | L | L |
| 6 | VL | VL |
| 7 | VVL | VVL |

## 7. RESULTS AND DISCUSSIONS

The proposed Fuzzy based Sliding Mode Control (FSMC) scheme is applied to the multi-machine power system. The controlled system is simulated using Matlab Simulink. The performance of proposed control scheme i.e. Fuzzy based Sliding Mode Control Power System Stabilizer (FSMCPSS) is compared to the performance of a conventional Power System Stabilizer(CPSS), Fuzzy Power System Stabilizer (FPSS), Sliding Mode Control based Power System Stabilizer (SMCPSS) and with the system without Power System Stabilizer (NOPSS).

The different designs of PSS is compared in terms of bus voltages $(V_{B1}, V_{B2}, V_{B3})$, line power, difference between rotor angle deviation of machine 1 and machine 2 $(dtheta_1 - dtheta_2)$, speed of the machines. $(\omega_1$ and $\omega_2)$, terminal voltages of machines

($Vt_1$ and $Vt_2$). Fig.10 shows the response of bus voltages ($V_{B1}$, $V_{B2}$, $V_{B3}$) with NOPSS, CPSS, FPSS, SMCPSS, and FSMCPSS.

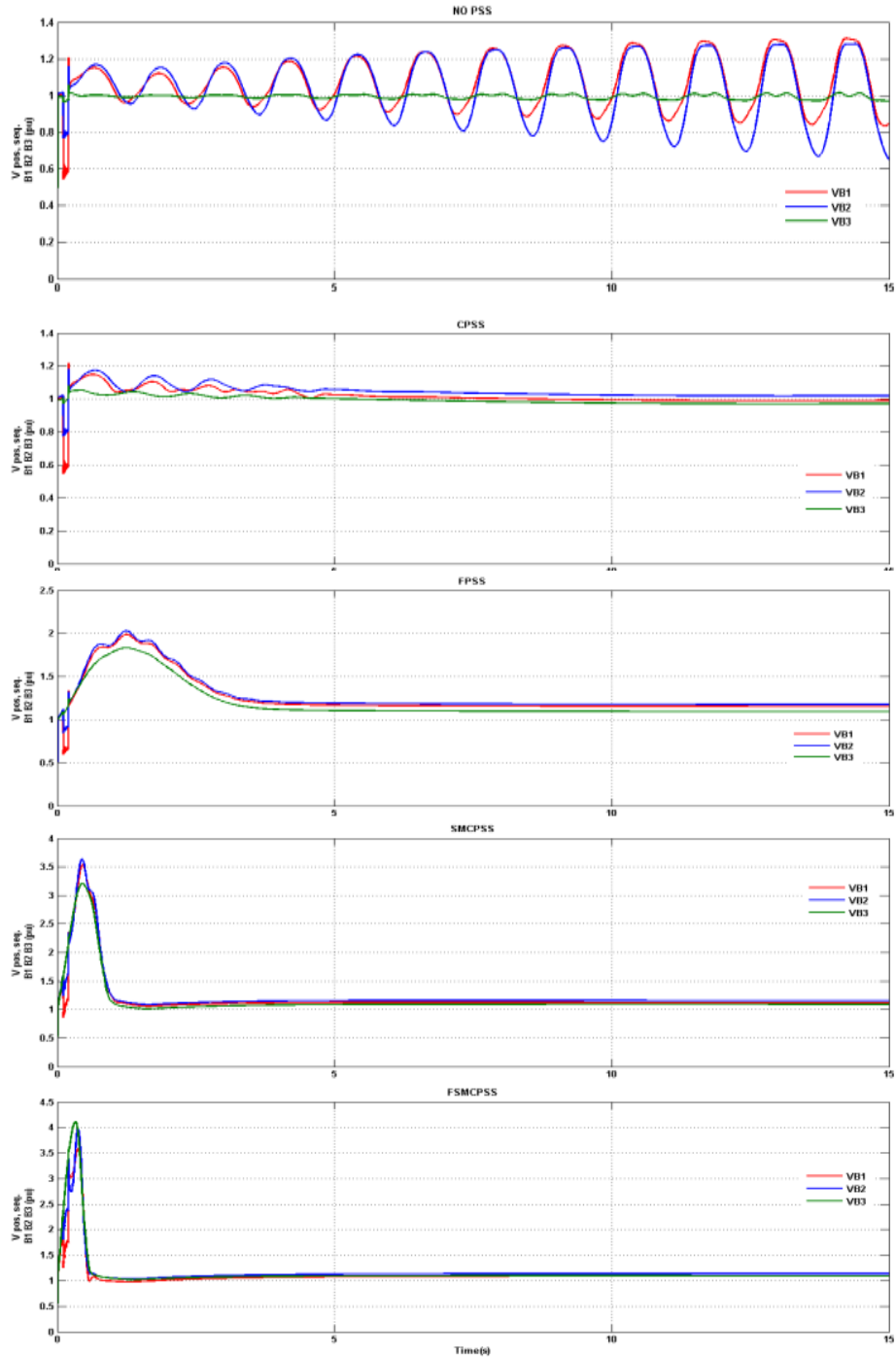

Fig.10. Response of bus voltages ($V_{B1}$, $V_{B2}$, $V_{B3}$)

Response of bus voltages ($V_{B1}, V_{B2}, V_{B3}$) with FSMCPSS gives better result. Settling time is very minimal in FSMCPSS (around 1sec) but the initial overshoot is slightly on the higher side. Fig. 11 shows the response of Line power with NOPSS, CPSS, FPSS, SMCPSS, and FSMCPSS.

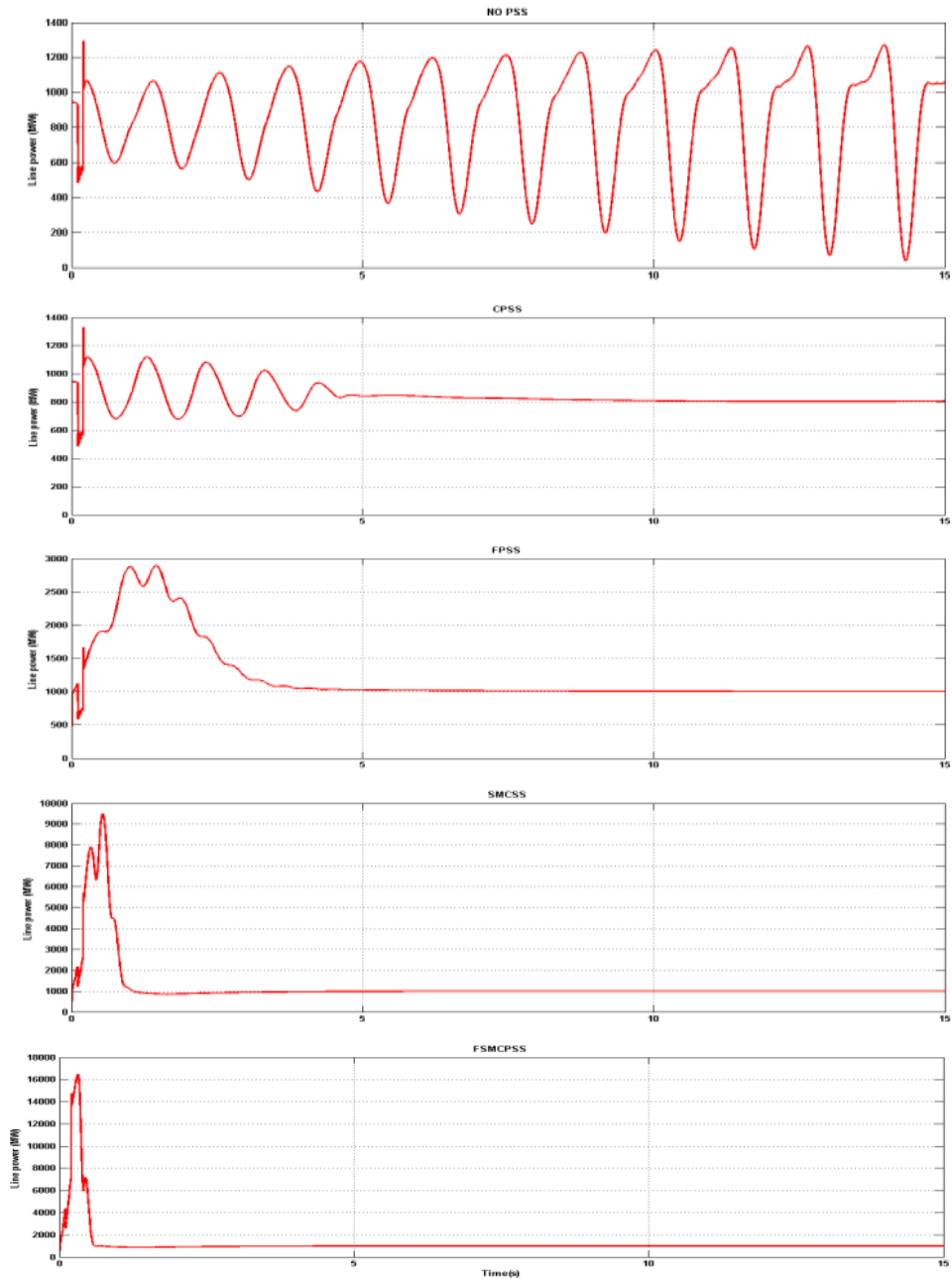

Fig.11. Response of Line Power

Response of Line power with FSMCPSS gives better result in terms of settling time and lesser oscillations. Fig.12 shows the response of difference between rotor angle deviation with NOPSS, CPSS, FPSS, SMCPSS, and FSMCPSS.

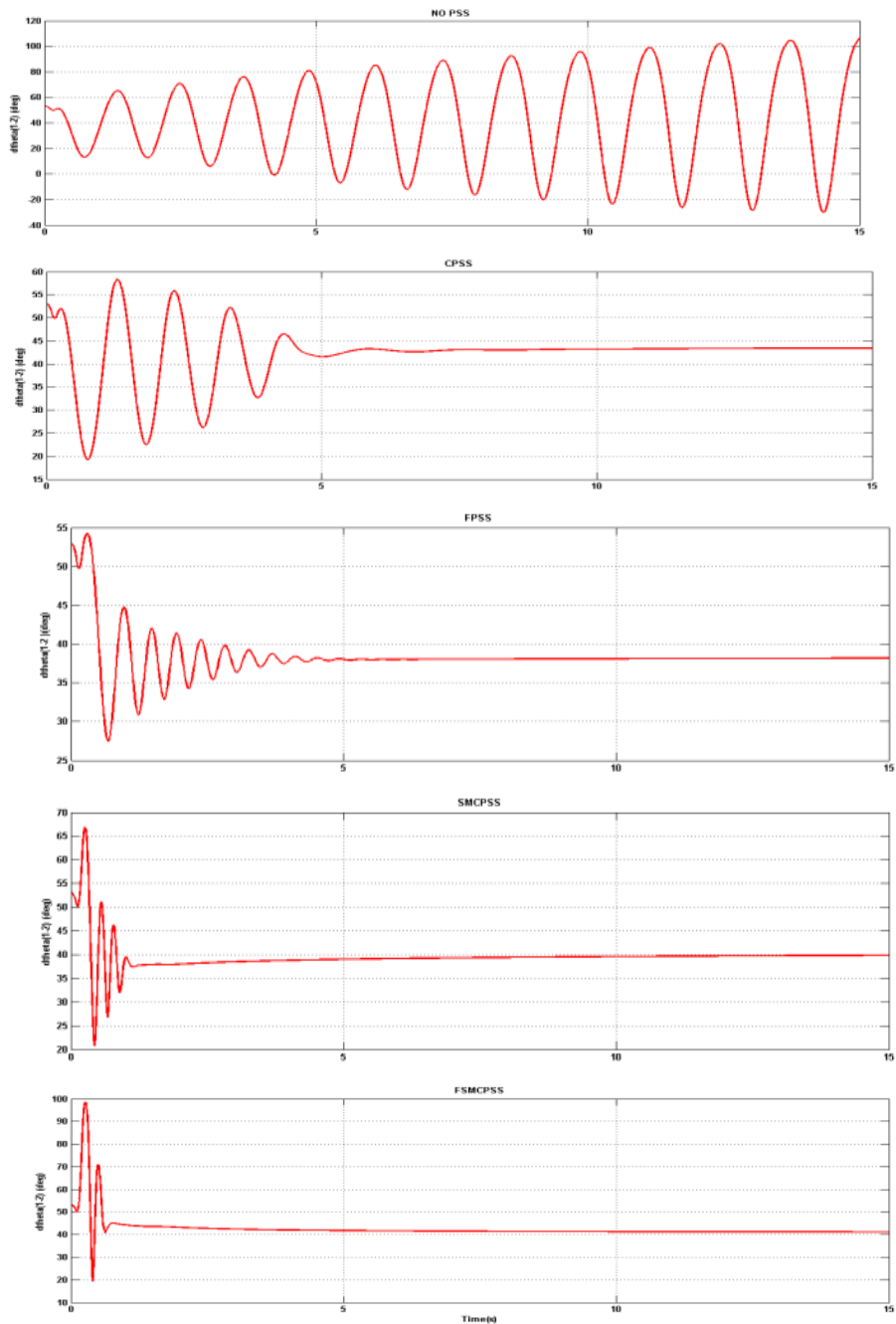

Fig.12. Response of difference between rotor angle deviations

The rotor angle deviation of system with FSMCPSS is within the specified limit from the initial start unlike other PSS which takes a longer time to settle down. Fig.13 shows the response of speed of the machine (ω1 and ω2) with NOPSS, CPSS, FPSS, SMCPSS, and FSMCPSS.

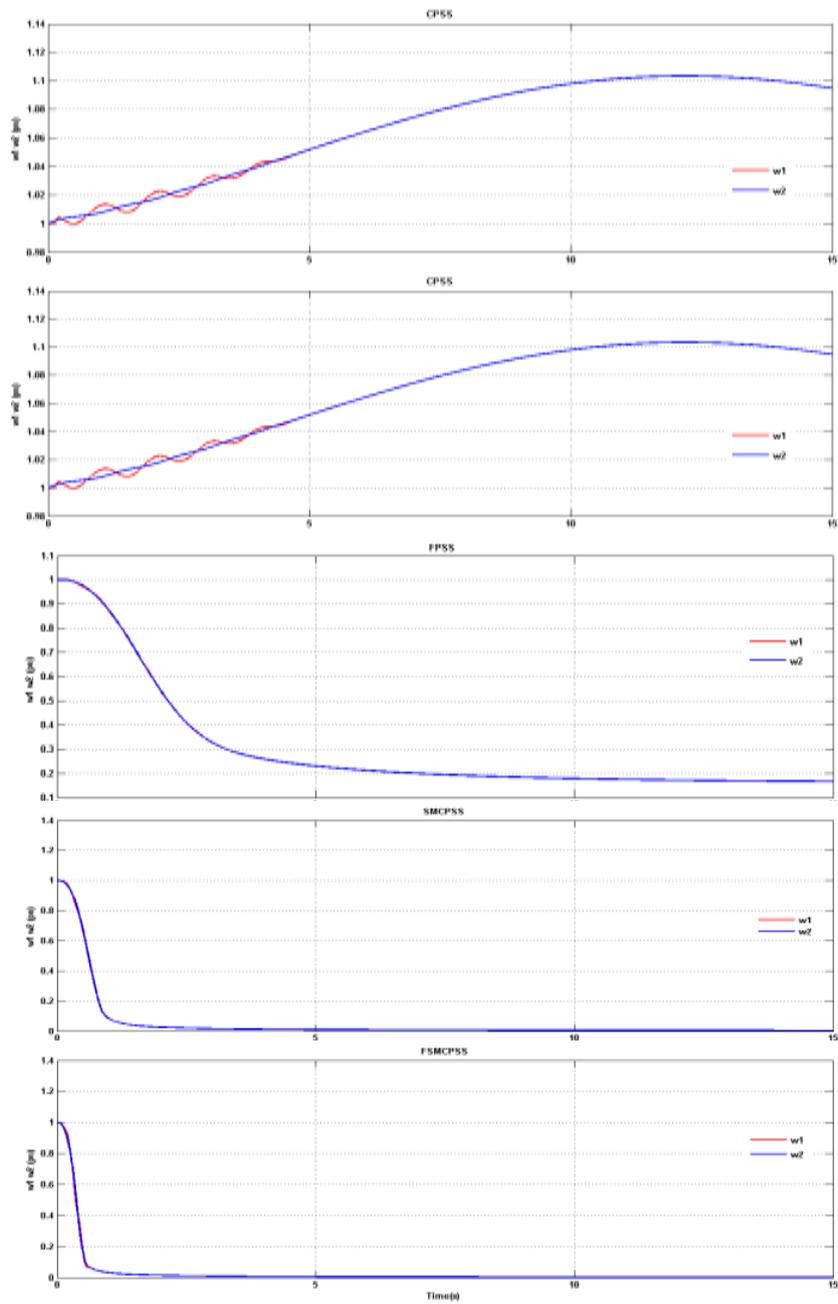

Fig.13. Response of speed of the machine (ω1 and ω2)

The speed of response of both the machines is very high in case of FSMCPSS and the machines attain the desired speed almost in 1.25 seconds. Fig.14 shows the response of terminal of the machine with NOPSS, CPSS, FPSS, SMCPSS, and FSMCPSS.

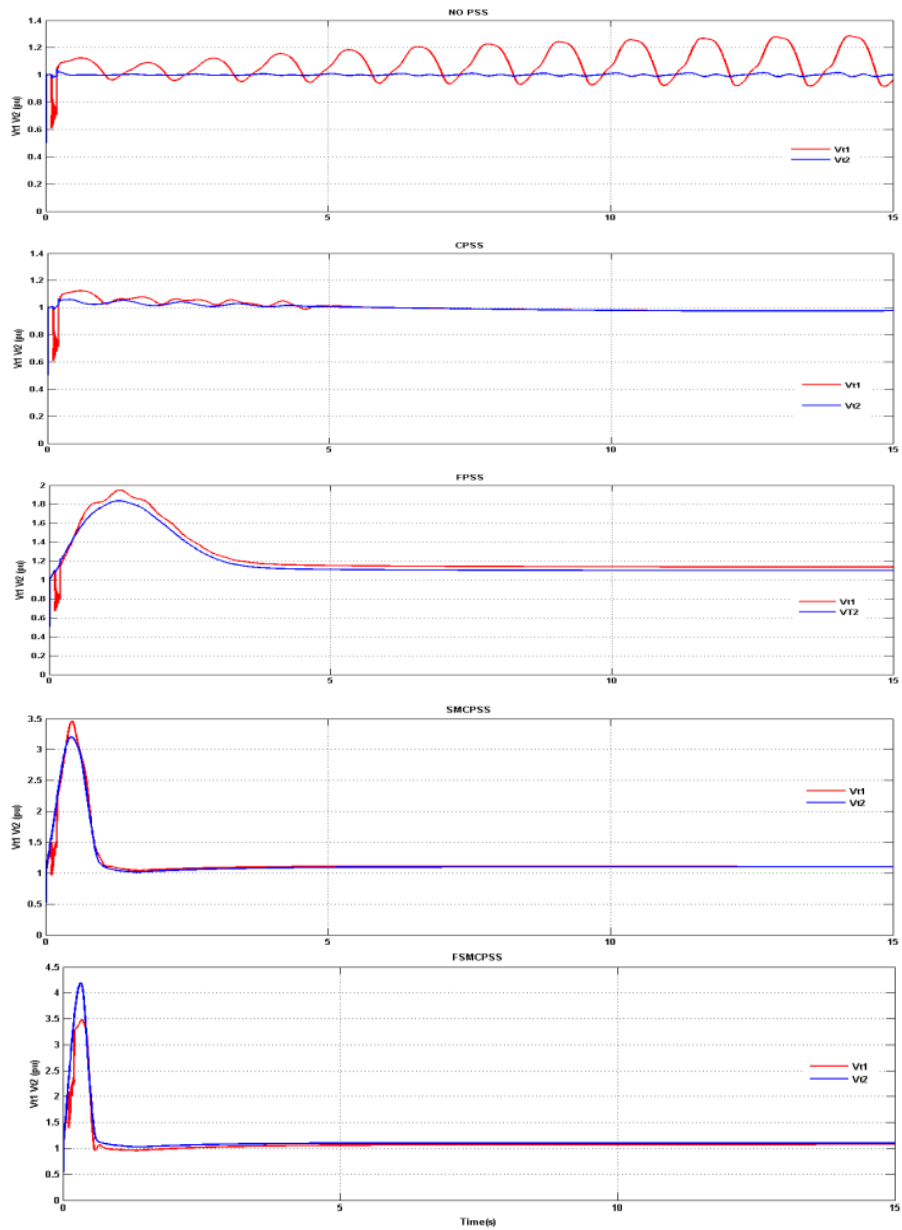

Fig.14. Response of terminal voltages of the machine

Response of terminal voltages of the machine with FSMCPSS gives better result in terms of settling time and lesser oscillations.

## 8. CONCLUSION

As power system is a highly complex system and the system equations are nonlinear as the parameters vary due to noise and load fluctuation, the Fuzzy based Sliding Mode Control based Power System Stabilizer enhances the stability of the system and improves the dynamic response of the system operating in faulty conditions in a better way and it has also effectively enhanced the damping of electromechanical oscillations. According to non-linear simulation results of a multi-machine power system, it is found that the Fuzzy based Sliding Mode Controller work well and is robust to change in parameters of the system and to disturbance acting on the system and also indicate that the proposed controller achieves a better performance than the Sliding Mode Control based Power System Stabilizer (SMCPSS), Fuzzy based Power System Stabilizer (FPSS) and Conventional Power System Stabilizer (CPSS).